\documentclass[aps,prb,floatfix,twocolumn,groupedaddress,showpacs,letterpaper,amsmath,amssymb]{revtex4-2}
\usepackage{amsfonts,mathrsfs,amsthm}
\usepackage{amsmath,amssymb,graphicx,hyperref}
\usepackage[table]{xcolor}
\usepackage{tikz,orcidlink}
\usepackage{multirow}
\usepackage{booktabs}
\usepackage{makecell}
\usepackage{colortbl}
\usepackage{array}
\usepackage{bm}
\usepackage{diagbox}
\hypersetup{pdfnewwindow=true, colorlinks=true, linkcolor=blue, anchorcolor=blue, citecolor=blue, filecolor=blue, menucolor=blue, urlcolor=blue}
\newcommand{\beq}{\begin{equation}}
\newcommand{\eeq}{\end{equation}}
\newcommand{\bea}{\begin{eqnarray}}
\newcommand{\eea}{\end{eqnarray}}

\newcommand{\nn}{\nonumber}
\newcommand{\down}{\downarrow}
\newcommand{\up}{\uparrow}

\newlength{\Oldarrayrulewidth}

\begin{document}

\title{Catalog of cubic, symmetry-protected, non-Fermi liquid,\\Kondo-type exchange models for doublet impurities
}

\author{Anna I.\ T\'oth\orcidlink{0000-0001-6503-850X}}\email{atoth2@ed.ac.uk}
\author{Andrew D.\ Huxley}

\affiliation{School of Physics and Astronomy, Centre for Science at Extreme Conditions, The University of Edinburgh, Edinburgh EH9 3FD, United Kingdom}

\date{\today}

\begin{abstract}
  To identify what types of non-Fermi liquid (NFL) behavior are most likely to occur in cubic metals due to doublet impurities, 
  we derive every cubic symmetry-allowed, NFL, Kondo-type exchange coupling that does not need accidental degeneracy for its realization.
  We find three distinct types of NFL behavior: two-channel Kondo (2CK) behavior for a non-Kramers doublet impurity coupled to local $\Gamma_8$ conduction electrons; topological Kondo physics for a Kramers doublet impurity and $\Gamma_4$ or $\Gamma_5$ conduction electrons; and lastly, spin-half impurity spin-$\frac 3 2$ conduction electron Kondo behavior for a Kramers doublet impurity and  $\Gamma_8$ conduction electrons. The first two critical behaviors are not straightforward to realize. In the first case, 2CK physics is not guaranteed, since cubic symmetry does not prevent an effective spatial
  anisotropy from exceeding the 2CK coupling, which restores a Fermi liquid behavior. % in diluted, cubic, heavy fermion systems with non-Kramers doublet impurities.
In the second case, the topological Kondo interaction is guaranteed to dominate, however, the spin degeneracy of the conduction electrons needs to be lifted e.g.\ by a magnetic field\textemdash so that they can be represented by
  $\Gamma_4$ or $\Gamma_5$ triplets\textemdash which then also lifts the degeneracy of the Kramers doublet.  We find that the spin-half impurity spin-$\frac 3 2$ conduction electron, NFL, Kondo behavior has the greatest chance of existing in diluted, cubic compounds. We compute the thermodynamics of the topological Kondo model using the numerical renormalization group, and discuss the thermodynamics of the spin-half impurity spin-$\frac 3 2$ conduction electron Kondo model. We also identify candidate materials where the corresponding NFL behaviors could be observed.
  %With the numerical renormalization group, we compute the finite-size excitation spectrum of every cubic, Kondo-type exchange model with doublet impurities, as well as the thermodynamic properties of the topological Kondo model. We discuss the thermodynamics of the spin-half impurity spin-$\frac 3 2$ conduction electron Kondo model, and identify candidate materials where the corresponding NFL behavior could be observed.
\end{abstract}

%\pacs{71.10.Hf, 71.27.+a, 75.20.Hr}

\maketitle

\section{Introduction}
Increasingly many materials are synthesized that exhibit anomalous metallic phases at
sufficiently low temperatures\textemdash which, in certain cases, extend also to high temperatures\textemdash that
cannot be described by Landau’s Fermi liquid theory, and are consequently called non-Fermi liquids (NFLs) or exotic/strange/bad metals.
The first encounter with NFL phenomena took place decades ago in heavy fermion materials \cite{Ott83,Seaman91} and in hole-doped, high-$T_c$, cuprate superconductors (SCs) \cite{Suzuki87,Gurvitch87,Giraldo-Gallo18}. Later it was observed e.g.\  
in doped, high-$T_c$, iron-pnictide SCs \cite{Liu08,Fang09,Kasahara10} and in stoichiometric, transition metal oxides such as VO$_2$ \cite{Allen93,Qazilbash06}, and 
in some ruthenates \cite{Kostic98,Khalifah01,Lee02} %RuO42- electron config: [Kr] 4d7 5s1
and iridates \cite{Nakatsuji06,Cao07},
in  pure, intermetallic and transition metal compounds \cite{Pfleiderer01,Steiner03,Brando08},  as well as %The oxyanion of iridium Ir2O76-; any salt containing this anion
%electron config of Ir: [Xe] 4f14 5d7 6s2
in other $d$- and $f$-electron systems \cite{Stewart} and elsewhere \cite{Dardel93,Moser98,Bockrath99,Jaoui22}.
%NFL phenomena discoveblack in three- and low-dimensional materials are abundant and diverse, as there are many forms of 
%deviations from Fermi liquid-like behavior. 
In heavy fermion systems, NFL behavior is manifest e.g.\ in 
the electronic specific heat (or Sommerfeld) coefficient and in the magnetic susceptibility, which show either
diverging/logarithmic or mild power-law $T$-dependencies, as well as  
in the electrical resistivity which goes as $\propto T^{\,\alpha}$, with $0<\alpha<2$, down to the lowest temperatures attainable 
\cite{Stewart}. Identifying the underlying, microscopic origins of the various types of NFL behavior,
in most cases, remains one of the most important unsolved challenges  in strongly correlated electron physics. One unifying feature
of all NFL many-body systems is that their low-energy states cannot be interpreted in terms of independent, additive, quasiparticle excitations.
So far, only a handful of solvable, microscopic models and concepts have been developed for the description of NFL physics \cite{Tomonaga50,Luttinger63,Lee18,Halperin20,Chowdhury22}. Impurity quantum criticality is one of them. Obtaining the
low-lying excitation spectrum of overscreened, multi-channel
Kondo models \cite{Nozieres80} around their NFL fixed points
%in a large but finite volume
is straightforward thanks to powerful methods that include the numerical renormalization group (NRG) \cite{Wilson75} and conformal field theory (CFT) \cite{Affleck}.
These fixed points can be relevant for the description of NFL behavior in some dilute heavy fermion systems. Most notably in Ref.\ \cite{Cox87}, Cox proposed the two-channel Kondo (2CK) model to describe U-based, cubic, heavy fermion systems where the U atoms have an $f^2$ electronic configuration with a $\Gamma_3$ non-Kramers  doublet ground state which hybridizes with $\Gamma_8$ conduction electrons. This work has been followed up by a large number of experimental attempts 
to observe the 2CK effect in numerous diluted, cubic and tetragonal, 4 and $5f$-electron systems
\cite{Seaman91,Aliev94,Amitsuka94,Onimaru16,Yamane18,Yanagisawa19}. In 2018, Y$_{1-x}$Pr$_x$Ir$_2$Zn$_{20}$ \cite{Yamane18,Yanagisawa19} emerged as the most accepted realization of the 2CK effect in a cubic, heavy fermion system\textemdash decades on from Cox's original proposal \cite{Cox87}. 
To find out why it is inherently difficult to realize 2CK physics in dilute heavy fermion systems, and if other types of NFL quantum impurity behavior are more likely in the same setting, we revisit the symmetry considerations that form the basis of Cox's work.  We identify every NFL state that can occur in cubic metals due to doublet impurities without accidental degeneracy by constructing all cubic symmetry-allowed, Kondo-type exchange interactions and solving them with NRG.

\section{Construction} To construct a NFL quantum impurity model, the necessary conditions are the degeneracy of the impurity states ($d_I\geq 2$), the degeneracy of the local conduction electron states ($d_c\geq3$), and to achieve overscreening: $d_I < d_c$.
%These degrees of freedom are coupled to each other in a form that is dictated by the symmetry of the crystal lattice.
These requirements can only be satisfied by the point groups $T,T_h,O,T_d,O_h$  (using the Sch\"onflies symbols for point groups) and the corresponding double groups having at least three-dimensional irreducible representations (irreps). The symmetry-based selection rules for 2CK models of U$^{4+}$ and Ce$^{3+}$ ions in metals written by Cox \cite{Cox92,Cox98} do not rule out tetragonal or hexagonal systems from hosting 2CK physics, however, as was shown in Ref.\ \cite{Toth11}, in a tetragonal system, only accidental degeneracy can lead to 2CK physics; similar considerations apply to the hexagonal case. Note that the same accidental degeneracy requirement arises with  impurity models in which different screening channels correspond to different irreps of the point group as in Ref.\ \cite{Patri20}, where, in a cubic crystal field, it was found that a novel NFL fixed point emerges due to the screening of a $\Gamma_3$ non-Kramers doublet impurity simultaneously by $\Gamma_8$ and $\Gamma_6$ electrons with Kramers degeneracy. However, in such a setting, the equivalence of the $\Gamma_8$ and $\Gamma_6$ conduction electrons cannot be ensured and it can only be the result of accidental degeneracy, as the point group symmetry allows for the presence of a crystal field that splits the two  multiplets breaking the channels symmetry necessary for reaching the novel NFL fixed point.

Here we discuss the case of $d_I=2$ which might be the simplest one experimentally but the following analysis can be easily repeated for $d_I=3$. We concentrate on the octahedral or cubic point group $O$ (which is isomorphic to $T_d$) as it covers every NFL state that can arise due to its identical Clebsch--Gordan coefficients with the other possible symmetries \cite{Koster63}. Looking at the character table of $O$, we find that four cases satisfy the above criteria that do not require additional accidental symmetry to achieve a NFL state. They are summarized in Table \ref{tab:summary_of_cases}.
%\vspace{-.2cm}
%\setlength\arrayrulewidth{1.1pt}
%\newcolumntype{?}{!{\vrule width 1.2pt}}
%\renewcommand{\arraystretch}{2.7}
%\begin{table}[hbt]
%\includegraphics[width=1\linewidth]{tablazat2.png}
%\newcolumntype{?}{!{\vrule width 1.2pt}}
\newcolumntype{?}{!{\vrule width 1.2pt}}
\renewcommand{\arraystretch}{2.7}
\begin{table}[hbt]
  \begin{tabular}{|c|c|cc|}
    \cline{3-4}
    \multicolumn{2}{c}{}& \multicolumn{2}{|c|}{\makecell{Local conduction electron\\}}\\
    \cline{3-4}
    \multicolumn{2}{c}{}& \multicolumn{1}{|c}{\makecell{$\Gamma_{4/5}$ $\left(T_{1/2}\right)$\\triplet}}&\multicolumn{1}{|c|}{\makecell{$\Gamma_{8}$ quartet}}\\
    \Xhline{\arrayrulewidth}
    \multicolumn{1}{|c}{}&\multicolumn{1}{|c}{\makecell{non-Kramers\\$\Gamma_3$ $(E)$ doublet}}&\multicolumn{1}{|c}{\makecell{Fermi liquid\\(FL)}}&\multicolumn{1}{|c|}{\diagbox[height=1.25cm,width=2cm,innerwidth=2.52cm,innerleftsep=.1cm,innerrightsep=.252cm]{\makecell{2CK\\NFL}}{\makecell{FL\\\quad}}}\\[5pt]
    \cline{2-4}
    \multicolumn{1}{|c}{\multirow{-2}{*}{Impurity}}&\multicolumn{1}{|c}{\makecell{Kramers\\$\Gamma_{6/7}$ doublet}}&\multicolumn{1}{|c}{\makecell{Topological\\Kondo\\ NFL}}&\multicolumn{1}{|c|}{\makecell{Impurity spin$-\frac 1 2$\\conduction $e^-$$-\frac 3 2$\\NFL}}\\
    \Xhline{\arrayrulewidth}
  \end{tabular}
%  \includegraphics{tablazat2}
%  \begin{tabu} to \textwidth {|[1.2pt] c| c|[1.2pt] | cc p{2.6cm}|[1.2pt]}
%    \tabucline[1.2pt]{3-4}
%    \multicolumn{2}{c}{}&\multicolumn{2}{|[1.2pt]c|[1.2pt]}{\makecell{Local conduction electron}}\\
%    \tabucline[0.4pt]{3-4}
%    \multicolumn{2}{c}{}&\multicolumn{1}{|[1.2pt]c}{\makecell{$\Gamma_{4/5}$ $\left(T_{1/2}\right)$\\triplet}}&\multicolumn{1}{|c|[1.2pt]}{\makecell{$\Gamma_{8}$ quartet}}\\
%    \tabucline[1.2pt]{1-4}
%    \multicolumn{1}{|[1.2pt]c}{}&\multicolumn{1}{|c}{\makecell{non-Kramers\\$\Gamma_3$ $(E)$ doublet}}&\multicolumn{1}{|[1.2pt]c}{\makecell{Fermi liquid\\(FL)}}&\multicolumn{1}{|c|[1.2pt]}{
%      \diagbox[linewidth=0.4pt,height=1.4cm,innerwidth=3cm,innerleftsep=.13cm,innerrightsep=.13cm]{\makecell{2CK\\NFL}}{\makecell{FL\\\quad}}}\\[.7pt]
%    \cline{2-4}
%    \multicolumn{1}{|[1.2pt]c}{\multirow{-2}{*}{Impurity}}&\multicolumn{1}{|c}{\makecell{Kramers\\$\Gamma_{6/7}$ doublet}}&\multicolumn{1}{|[1.2pt]c}{\makecell{Topological\\Kondo\\ NFL}}&\multicolumn{1}{|c|[1.2pt]}{\makecell{Impurity spin-half\\conduction $e^-$$-\frac 3 2$\\NFL}}\\
%    \tabucline[1.2pt]{1-4}
%  \end{tabu}
  \caption{The four cases that satisfy the necessary conditions for the construction of a cubic, NFL, quantum impurity Hamiltonian. Throughout the text, we use the notations of Ref.\ \cite{Koster63} when referring to irreps. We also listed the Mulliken symbols for the irreps of $O/T_d$ in parentheses. Note that Mulliken symbols are not in use for the associated double groups $O^\prime/T_d^\prime$.}
  \label{tab:summary_of_cases}
\end{table}
\renewcommand{\arraystretch}{1}
The construction of the exchange couplings goes by creating every possible irreducible electron-hole tensor operator from the local conduction electrons, as well as creating every possible irreducible tensor operator from the impurity ket and bra states, and then taking the scalar product of the irreducible tensor operators that transform the same way under cubic symmetry. The resulting cubic scalars are the only forms of exchange coupling that can appear in an exchange Hamiltonian.  

The first and simplest case with the least number of degrees of freedom is when, in a cubic metal, the impurity ground state is a $\Gamma_3$
non-Kramers doublet irrep and it hybridizes with $\Gamma_4$ or $\Gamma_5$ conduction electrons. However it does not result in NFL behavior.
For this reason, we relegated the derivation of the exchange Hamiltonian to App.\ \ref{app:G3_G5}.
The second case is Cox's construction of the 2CK model in cubic symmetry. 
\section{Cox's 2CK Hamiltonian} To construct a tractable model, valid at low temperatures, Cox assumed the impurity to be in a time-reversal invariant,
$f^2$-state
whose ground state transforms as a $\Gamma_3$ non-Karmers doublet irrep of the group $O$ \cite{Cox87}.
%and surrounding $\Gamma_8$ conduction electrons \cite{Cox87}.
%Here $\,\mathscr{T}\,=\,\{{\cal I},\,{\cal T}\}\,$ is the group of time-reversal,
%$\,{\cal I}\,$ the identity,
%$\,{\cal T}\,$  the time-reversal operator.
%$T_d$ of all proper rotations sending a regular tetrahedron into itself, 
The doublet impurity ground state was specified in Ref.\ \cite{Cox87,Satten60} in terms of the eigenvectors, $\,|J_z\rangle\,$, of the operator $\,{\hat J}_z\,$
in the $\,J\,=\,4\,$ multiplet, with the quantization axis chosen parallel to the $c$-axis of
the crystal as
$\,|\Gamma_{3}^{\,3z^2-r^2}\rangle\equiv\sqrt{\frac{5}{12}}\,|0\rangle\,-\,\sqrt{\frac{7}{24}}\,\left(| 4\rangle\,+\,| -4\rangle\right)\,$
and $\,|\Gamma_{3}^{\sqrt{3}(x^2-y^2)}\rangle\equiv\frac{1}{\sqrt{2}}\left(|2\rangle\,+\,| -2\rangle\right)\,$.
Note that the conclusions about the form of the exchange Hamiltonian do not depend on these specific forms, only on the transformation properties of the irreps. We only specified them to see how the exchange Hamiltonian transforms under time-reversal.
Cox further assumed that these localized $f$-levels would hybridize primarily with  $l\,=\,3,J\,=\,\frac 5 2$ conduction electrons transforming as a Kramers quartet, i.e.\ the $\Gamma_8$ irrep of $O^\prime$.  They are created by the cubic, irreducible tensor operator \cite{Cox87,Pappalardo61}
\bea
{\bm \Psi^{\Gamma_8\,\dagger}}\equiv\left(
\renewcommand{\arraystretch}{1.35}
\begin{array}{c}
\Psi_{\frac 3 2}^{\Gamma_8\,\dagger}\\
\Psi_{\frac 1 2}^{\Gamma_8\,\dagger}\\
\Psi_{-\frac 1 2}^{\Gamma_8\,\dagger}\\
\Psi_{-\frac 3 2}^{\Gamma_8\,\dagger}
\end{array}
\renewcommand{\arraystretch}{1}
\right)\equiv
\left(
\begin{array}{c}
\sqrt{\frac 5 6}\,\psi_{\frac 5 2}^\dagger\,+\,\frac 1{\sqrt 6}\,\psi_{-\frac 3 2}^\dagger\\
\psi_{\frac 1 2}^{\dagger}\\
\psi_{-\frac 1 2}^{\dagger}\\
\sqrt{\frac 5 6}\,\psi_{-\frac 5 2}^{\dagger}\,+\,\frac 1{\sqrt 6}\,\psi_{\frac 3 2}^{\dagger}
\end{array}
\right)
\eea
with $\psi_{J_z}^{\dagger}\,\equiv\,\psi_{J_z}^{l=3\,J=\frac 5 2\,\dagger}$.
%Here again, the expressions in terms of $\,|J_z\rangle\,$ could be dropped, we only kept them to make the role of time-reversal symmertry more apparent.
With this choice, the quartet components are connected by time-reversal as
$\,{\cal T}\,\Psi_{\,\left|J_z\right|}^{\Gamma_8\,\,\dagger}\,{\cal T}^{-1}=\,-\,\Psi_{\,-\left|J_z\right|}^{\Gamma_8\,\dagger}\,$, with $\,{\cal T}\,$ the time-reversal operator \footnote{The signs can be checked by acting on the one-electron wavefunction with the time-reversal operator, $-\,i\sigma_y\,{\mathcal K}$ where ${\mathcal K}$ is the complex conjugation.}. Its adjoint quartet, ${\bm \Psi^{\Gamma_8}}$, made up of
single-electron annihilation operators, can be defined as \cite{Messiah}
\bea
{\bm \Psi_{}^{\Gamma_8}}
%\left(
%\renewcommand{\arraystretch}{1.35}
%\begin{array}{c}
%\Psi^{\Gamma_8}_{\frac 3 2}\\
%\Psi^{\Gamma_8}_{\frac 1 2}\\
%\Psi^{\Gamma_8}_{-\frac 1 2}\\
%\Psi^{\Gamma_8}_{-\frac 3 2}
%\end{array}
%\renewcommand{\arraystretch}{1}
%\right)
\equiv\left(
\renewcommand{\arraystretch}{1.35}
\begin{array}{c}
\left({\Psi_{-\frac 3 2}^{\Gamma_8\,\dagger}}\right)^\dagger\\
-\left({\Psi_{-\frac 1 2}^{\Gamma_8\,\dagger}}\right)^\dagger\\
\left({\Psi_{\frac 1 2}^{\Gamma_8\,\dagger}}\right)^\dagger\\
-\left({\Psi_{\frac 3 2}^{\Gamma_8\,\dagger}}\right)^\dagger
\end{array}
\renewcommand{\arraystretch}{1}
\right)=
\left(
\begin{array}{c}
\sqrt{\frac 5 6}\psi_{-\frac 5 2}+\frac 1{\sqrt 6}\psi_{\frac 3 2}\\
-\psi_{-\frac 1 2}\\
\psi_{\frac 1 2}\\
-\sqrt{\frac 5 6}\psi_{\frac 5 2}-\frac 1{\sqrt 6}\psi_{-\frac 3 2}
\end{array}
\right)\,.
\eea
%Note that the components of the quartet are not connected by time-reversal symmetry the same way as for its adjoint.
%It transforms under time-reversal as $\,{\cal T}\,\Psi^{\,\left|J_z\right|}_{\Gamma_{8}}\,{\cal T}^{-1}=\,-\,\Psi^{\,-\left|J_z\right|}_{\Gamma_8}\,$. 
To build an exchange Hamiltonian, the impurity ket and bra vectors are combined according to 
$\Gamma_3\,\otimes\,\Gamma_3\,=\,\Gamma_1\,\oplus\,\Gamma_2\,\oplus\,\Gamma_3$, % to have particle number conserving operators,
whereas the product of the spinor irreps
can be decomposed as follows \cite{Koster63}:
\bea
\Gamma_8\,\otimes\,\Gamma_8\,=\,\Gamma_1\,\oplus\,\Gamma_2\,\oplus\,\Gamma_3\,\oplus\,2\Gamma_4\,\oplus\,2\Gamma_5\,.\label{eq:G8xG8}
\eea
So there are three independent terms in the exchange Hamiltonian,
as only the scalar products of identical irreps create cubic invariants.
The first term 
\bea
    {\cal H}^{\Gamma_1\otimes\Gamma_1}_{\textrm{\,Cox}}={\cal J}^{\Gamma_1\otimes\Gamma_1}_{\textrm{\,Cox}}\sum_{\substack{\delta\in\left\{3z^2-r^2,\right.\\\left.\sqrt{3}(x^2-y^2)\right\}}}|\Gamma_{\,3}^{\,\delta}\rangle\langle\Gamma_{\,3}^{\,\delta}|\sum_{\mu=-3/2}^{3/2}n^{\,\Gamma_8}_{\,\mu}\nn\\
    %    =
    %\left(n_{3z^2-r^2}+n_{\sqrt{3}(x^2-y^2)}\right)\left(n_{\frac 3 2}+n_{\frac 1 2}+n_{-\frac 1 2}+n_{-\frac 3 2}\right)
\eea
with $n^{\,\Gamma_8}_{\,\mu}\,\equiv\,(-)^{\mu\,}\Psi_{\mu}^{\Gamma_8\,\dagger}\Psi^{\Gamma_8}_{-\mu}$, which describes potential scattering and is diagonal in the impurity and local conduction electron degrees of freedom.  It is exactly marginal around the one- and two-channel Kondo fixed points, so its presence does not affect our conclusions. The second and third cubic scalars are found to be
\bea
{\cal H}^{\Gamma_2\otimes\Gamma_2}_{\textrm{\,Cox}}&=&{\cal J}^{\Gamma_2\otimes\Gamma_2}_{\textrm{\,Cox}}\,\left(|\Gamma_{\,3}^{\,\alpha}\rangle\langle\Gamma_{\,3}^{\,\beta}|\,-\,
|\Gamma_{\,3}^{\,\beta}\rangle\langle\Gamma_{\,3}^{\,\alpha}|\right)\nn\\
&&\times\,\left(\Psi_{-\frac 3 2}^{\,\Gamma_8\,\dagger}\Psi^{\,\Gamma_8}_{\frac 1 2}\,-\,\Psi_{-\frac 1 2}^{\,\Gamma_8\,\dagger}\Psi^{\,\Gamma_8}_{\frac 3 2}\,-\,h.c.\right)\,,\\
{\cal H}^{\Gamma_3\otimes\Gamma_3}_{\textrm{\,Cox}}&=&{\cal J}^{\Gamma_3\otimes\Gamma_3}_{\textrm{\,Cox}}\,\left[\left(|\Gamma_{3}^{\beta}\rangle\langle\Gamma_{3}^{\beta}|\,-\,
|\Gamma_{3}^{\alpha}\rangle\langle\Gamma_{3}^{\alpha}|\right)\right.\nn\\
&&\times\left(n^{\,\Gamma_8}_{-\frac 3 2}\,-\,n^{\,\Gamma_8}_{-\frac 1 2}\,-\,n^{\,\Gamma_8}_{\frac 1 2}\,+\,n^{\,\Gamma_8}_{\frac 3 2}\right)\,\nn\\
&&+\,\left(|\Gamma_{\,3}^{\,\alpha}\rangle\langle\Gamma_{\,3}^{\,\beta}|\,+\,
|\Gamma_{\,3}^{\,\beta}\rangle\langle\Gamma_{\,3}^{\,\alpha}|\right)\nn\\
&&\times\left.\left(\Psi_{-\frac 3 2}^{\,\Gamma_8\,\dagger}\Psi^{\,\Gamma_8}_{\frac 1 2}\,+\,\Psi_{-\frac 1 2}^{\,\Gamma_8\,\dagger}\Psi^{\,\Gamma_8}_{\frac 3 2}\,+\,h.c.\right)\right]\,
\eea
with state labels  $\alpha\equiv3z^2-r^2,\beta\equiv\sqrt{3}(x^2-y^2)$. With the following identifications, % $S^+\equiv |\Gamma_3^{\sqrt{3}(x^2-y^2)}\rangle\langle \Gamma_3^{3z^2-r^2}|\,,$
\bea
S^{\textrm{\,imp}}_{+}&\equiv&|\Gamma_3^{\sqrt{3}(x^2-y^2)}\rangle\langle \Gamma_3^{3z^2-r^2}|\,,\quad S^{\textrm{\,imp}}_-={S^{\textrm{\,imp}}_+}^\dagger\nn\\
S^{\textrm{\,imp}}_{z}&\equiv&\frac 1 2\left(|\Gamma_3^{\sqrt{3}(x^2-y^2)}\rangle\langle \Gamma_3^{\sqrt{3}(x^2-y^2)}|-|\Gamma_3^{3z^2-r^2}\rangle\langle \Gamma_3^{3z^2-r^2}|\right)\nn\\
s^1_{+}&\equiv&\Psi_{\frac 3 2}^{\,\Gamma_8\,\dagger}\Psi^{\,\Gamma_8}_{-\frac 1 2}\,,\,\,\, s^1_z\equiv\frac 1 2\left(n^{\,\Gamma_8}_{\frac 3 2}\,-\,n^{\,\Gamma_8}_{-\frac 1 2}\right)\,,
\,\,\, s^1_-\equiv s^{1\,\dagger}_+\nn\\
s^2_+&\equiv&\Psi_{-\frac 3 2}^{\,\Gamma_8\,\dagger}\Psi^{\Gamma_8}_{\frac 1 2}\,,\,\,\, s^2_z\equiv\frac 1 2\left(n^{\,\Gamma_8}_{-\frac 3 2}\,-\,n^{\,\Gamma_8}_{\frac 1 2}\right)\,,
\,\,\, s^2_-\equiv s^{2\,\dagger}_+\nn\\
&&
\eea
where the vector operators, ${\vec S}^{\textrm{\,imp}},\,{\vec s^{\,1}},\,{\vec s^{\,2}}$ satisfy the canonical commutation relations, 
we can map ${\cal H}^{\Gamma_2\otimes\Gamma_2}_{\textrm{\,Cox}}+{\cal H}^{\Gamma_3\otimes\Gamma_3}_{\textrm{\,Cox}}$ onto the spatially anisotropic 2CK model
\bea
   &&{\cal H}^{\Gamma_2\otimes\Gamma_2}_{\textrm{\,Cox}}=4\,{\cal J}^{\Gamma_2\otimes\Gamma_2}_{\textrm{\,Cox}}\,S^{\textrm{\,imp}}_y\left(s_y^1\,+\,s_y^2\right)\,,\\
&&{\cal H}^{\Gamma_3\otimes\Gamma_3}_{\textrm{\,Cox}}=4\,{\cal J}^{\Gamma_3\otimes\Gamma_3}_{\textrm{\,Cox}}\left[S^{\textrm{\,imp}}_x\left(s_x^1\,+\,s_x^2\right)+S^{\textrm{\,imp}}_z\left(s_z^1\,+\,s_z^2\right)\right]\nn\\
&&
    \eea
    where the upper indices 1 and 2 distinguish the two channels.  Note that the indices $x,y,z$ do not refer to actual spatial directions they are only effective spatial indices. 
    The above derivation shows that cubic symmetry ensures equal couplings to the two channels. The effective spatial anisotropy is an irrelevant perturbation around the 2CK fixed point \cite{Emery92}. However, when ${\cal J}^{\Gamma_2\otimes\Gamma_2}_{\textrm{\,Cox}}>{\cal J}^{\Gamma_3\otimes\Gamma_3}_{\textrm{\,Cox}}$, spatial anisotropy is no longer a perturbation, it is the dominant term in the Hamiltonian and the system is a Fermi liquid at low temperatures. As the ratio of ${\cal J}^{\Gamma_2\otimes\Gamma_2}_{\textrm{\,Cox}}$ to ${\cal J}^{\Gamma_3\otimes\Gamma_3}_{\textrm{\,Cox}}$ is
not dictated by symmetry, the coupling of a $\Gamma_3$ impurity to local $\Gamma_8$ conduction electrons does not necessarily result in 2CK physics \footnote{The antiferromagnetic nature of the couplings is assumed to originate from an underlying Anderson Hamiltonian.}. 
\subsection{Properties under time-reversal}
While the $\Gamma_3$ non-Kramers doublet impurity states as well as ${\cal H}^{\Gamma_1\otimes\Gamma_1}_{\textrm{\,Cox}}$ are time-reversal invariants, 
${\cal H}^{\Gamma_2\otimes\Gamma_2}_{\textrm{\,Cox}}$ changes sign under time-reversal and ${\cal H}^{\Gamma_3\otimes\Gamma_3}_{\textrm{\,Cox}}$ breaks time-reversal symmetry.
\begin{figure}
\includegraphics[width=1\linewidth]{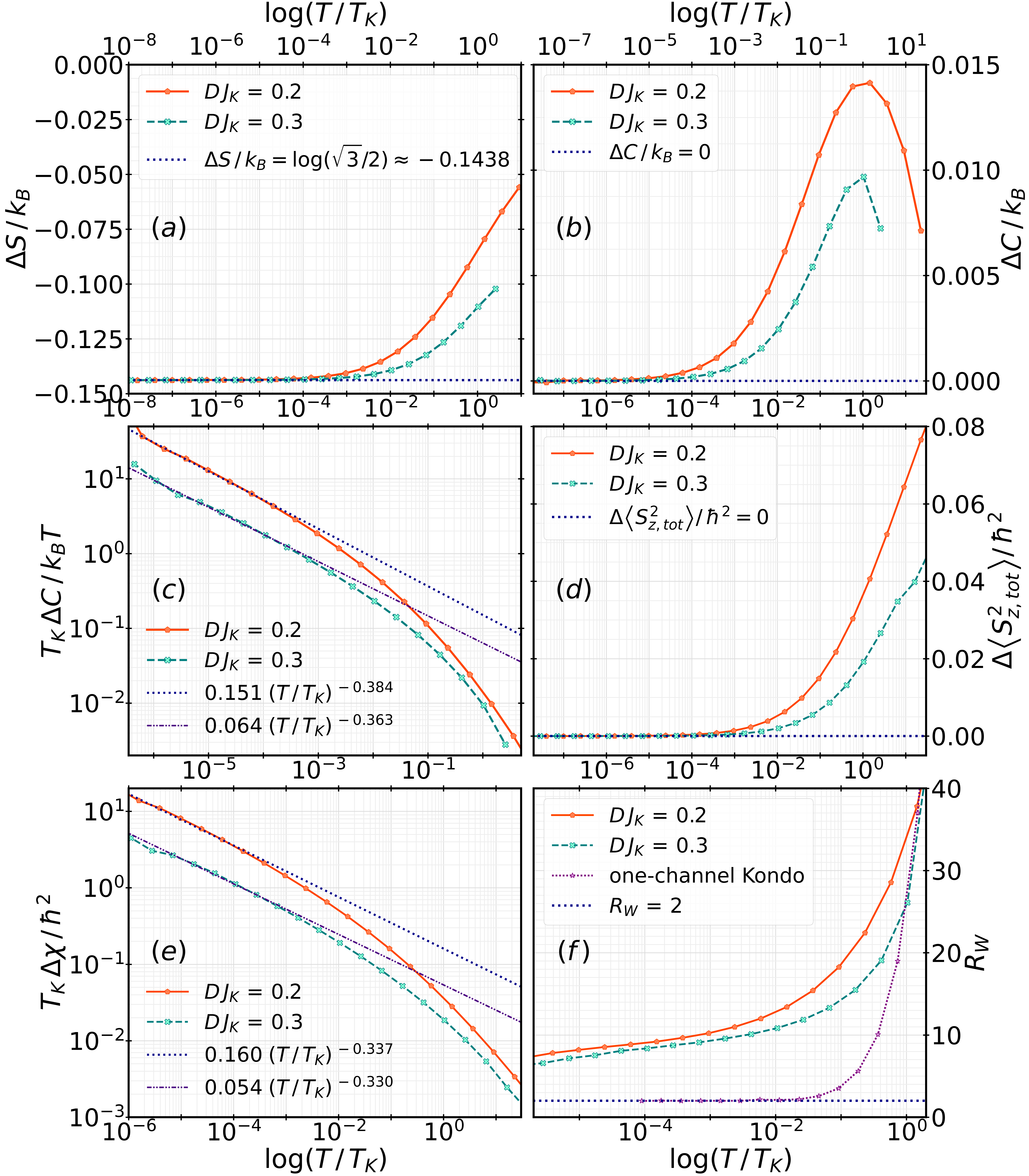}
\caption{Thermodynamics of the topological Kondo model with NRG. We made use of the SU(2)$_\textrm{spin}\times$U(1)$_\textrm{charge}$ symmetry of the 1.5CK model, and kept a minimum of 14k multiplets with the energy cutoff at its highest possible value and constant for each iteration step. We chose $\Lambda=2.5$ for the discretization parameter \cite{Wilson75}. For each curve, we used ten $z$ values for $z$-averaging \cite{Oliveira94}. The two values of the dimensionless Kondo coupling, ${\cal J}^{\Gamma_4\otimes\Gamma_4}_{\textrm{1.5CK}}$ appear in each panel in units of the bandwidth $D$. $(a)$ The impurity contribution to the entropy: at high temperatures, which is not captured well by NRG, the Kondo coupling is negligible, the impurity is a free spin and does not contribute to $S$. At low temperatures, below the Kondo scale, $T_K$, defined as the peak position in $\Delta C$,  the spin becomes overscreened with the expected $\log(\sqrt{3})$ contribution at $T=0$ to $S$. Our calculations reproduced the expected entropy difference of $\log(2)-\log(\sqrt{3})$ to four-digit precision. $(b)$ The impurity contribution to the specific heat. $T_K$ is defined as the location of the peak in the specific heat. $(c)$ The impurity contribution to the Sommerfeld coefficient. The expected $T^{-1/3}$ behavior is approximately reproduced over three decades. $(d)$ The impurity contribution to $\langle S_z^2\rangle$. $(e)$ The impurity contribution to $\chi$. The expected $T^{-1/3}$ behavior is approximately reproduced over three decades. $(f)$ The resulting Wilson ratio whose non-universality is reproduced, but the precision of the calculations is not high enough to see the leveling off as $T\to0$.   
}
\label{fig:1.5CK_TD}
\end{figure}
\section{Construction of the topological Kondo model in cubic symmetry}
The topological Kondo model\textemdash also called the spin-half impurity spin-one conduction electron Kondo model or the 1.5-channel Kondo (1.5CK) model \cite{Toth17}\textemdash, which has less degrees of freedom than the 2CK model and therefore could be considered simpler, was first studied by Fabrizio and Gogolin in Ref.\ \cite{Fabrizio94}. Newfound interest in it began with Ref.\ \cite{Beri12}, where it was suggested that the corresponding NFL Kondo effect could be realized with an impurity made of non-local Majorana zero modes coupled to spin-one conduction electrons in semiconductor nanowires, and that is where its name ``topological'' comes from. However it can also be described without referring to Majorana zero modes as the dominant coupling between a $\Gamma_6$ or $\Gamma_7$ Kramers doublet (odd-electron) impurity ground state and surrounding $\Gamma_4$ or $\Gamma_5$ conduction electrons \footnote{Note that the construction and the final form of the Hamiltonian is the same for both types of impurities as well as for both types of conduction electrons. For notational convenience, we refer to a $\Gamma_6$ impurity and $\Gamma_5$ conduction electrons in the text.}\cite{Toth17}. The construction goes the same way as in Cox's case with the representation indices replaced by $\Gamma_6$ for the impurity and $\Gamma_5$ for the conduction electrons. From the two cubic, $\Gamma_6$ spinors
$\left(
\begin{array}{c}
f^\dagger_{\up}\\
f^\dagger_{\down}
\end{array}
\right),\,
\left(
\begin{array}{c}
-\,f^{}_{\down}\\
f^{}_{\up}
\end{array}
\right),$
with $f^\dagger_{\mu}$ creating a spin-half impurity with orientation $\mu$, and normalized to anticommute, based on the tensor product decomposition
\bea
\Gamma_6\,\otimes\,\Gamma_6\,=\Gamma_1\,\oplus\,\Gamma_4\,,\label{eq:G6xG6}
\eea
we can form the following two irreducible, electron-hole, tensor operators
\bea\label{eq:spin_def}
n\equiv\sum_{\mu\in\{\uparrow,\downarrow\}}f^\dagger_{\mu}f^{}_{\mu},\,\,{\vec S}^{\textrm{\,imp}}\equiv
\frac{1}{2}\left[
\begin{array}{c}
f^\dagger_{\down}f^{}_{\up}+f^\dagger_{\up}f^{}_{\down}\\
i\left(f^\dagger_{\down}f^{}_{\up}-f^\dagger_{\up}f^{}_{\down}\right)\\
f^\dagger_{\up}f^{}_{\up}-f^\dagger_{\down}f^{}_{\down}
\end{array}
\right].
\eea
The product of the two $\Gamma_5$
conduction electron and hole tensor operators,
$\,
\left(
\begin{array}{c}
c^\dagger_{yz}\\
c^\dagger_{zx}\\
c^\dagger_{xy}\\
\end{array}
\right)$ and
$\left(
\begin{array}{c}
c_{yz}^{}\\
c_{zx}^{}\\
c_{xy}^{}
\end{array}
\right),\,$
whose spin degeneracies are lifted, can be decomposed according to 
$\Gamma_5\,\otimes\,\Gamma_5\,=\,\Gamma_1\,\oplus\,\Gamma_3\,\oplus\,\Gamma_4\,\oplus\,\Gamma_5$. Thus, apart from the ubiquitous and exactly marginal potential scattering term, there is only one cubic symmetry-allowed term in the exchange Hamiltonian, namely the topological Kondo model
\bea
&&{\cal H}^{\,\Gamma_4\otimes\Gamma_4}_{\textrm{\,1.5CK}}=\frac{i}{\sqrt{2}}\,{\cal J}^{\Gamma_4\otimes\Gamma_4}_{\textrm{1.5CK}}
    \left[\,S^{\textrm{\,imp}}_x\left(c^\dagger_{zx}c^{}_{xy}-\,c^\dagger_{xy}c^{}_{zx}\right)\right.\nn\\
      &&\left.+\,S^{\textrm{\,imp}}_y\left(c^\dagger_{xy}c^{}_{yz}-\,c^\dagger_{yz}c^{}_{xy}\right)+
      S^{\textrm{\,imp}}_z\left(c^\dagger_{yz}c^{}_{zx}-\,c^\dagger_{zx}c^{}_{yz}\right)\right]\nn\\
    &&\label{eq:top_Kondo1}
\eea
which can be further simplified as
\bea
    {\cal H}^{\Gamma_4\otimes\Gamma_4}_{\textrm{\,1.5CK}}={\cal J}^{\Gamma_4\otimes\Gamma_4}_{\textrm{1.5CK}}&&\left[\frac{1}{\sqrt{2}}\,S^{\textrm{\,imp}}_+\left(c^\dagger_0c^{}_{1}+c^\dagger_{-1}c^{}_0\right)+h.c.\right.\nn\\
     && \left.+\,S^{\textrm{\,imp}}_z\left(c^\dagger_1c^{}_{1}-c^\dagger_{-1}c^{}_{-1}\right)\right]\label{eq:top_Kondo2}
\eea
with the introduction of the spin-one tensor i.e.\ vector operator
\bea
\left(
\begin{array}{c}
c_{1}^\dagger\\
c_{0}^\dagger\\
c_{-1}^{\dagger}
\end{array}
\right)
\equiv
\left(
\begin{array}{c}
\frac{1}{\sqrt 2}\left(c_{yz}^\dagger+ic_{zx}^\dagger\right)\\
c_{xy}^\dagger\\
\frac{1}{\sqrt 2}\left(-c_{yz}^\dagger+ic_{zx}^\dagger\right)
\end{array}
\right).
\eea
This model is not only a cubic but also a spherical invariant since it is the scalar product of the impurity spin vector operator and an effective angular momentum operator for the conduction electrons \footnote{The generators of the continuous rotations are ${\vec S}^{\textrm{\,imp}}+{\vec L}^{\textrm{\,eff}}_{\textrm{cond}-e}$ with $\left(-c^\dagger_{\,1}c^{}_{\,0} - c^\dagger_{\,0}c^{}_{\,-1}, c^\dagger_{\,1}c^{}_1 - c^\dagger_{\,-1}c^{}_{\,-1}, c^\dagger_{\,-1}c^{}_{\,0} + c^\dagger_{\,0}c^{}_{\,1}\right) $  the $\left(L^\textrm{\,eff}_{1}, L^\textrm{\,eff}_{0}, L^\textrm{\,eff}_{-1}\right)$ vector operator components}. As there are no other cubic
symmetry-allowed terms in the exchange Hamiltonian, this term gives the dominant NFL behavior at low temperatures.
\subsection{Thermodynamics\\of the topological Kondo model}
The thermodynamics (TD) of the topological Kondo model were calculated in Ref.\ \cite{Fabrizio94}, where it was argued that this model has the same impurity contribution to the zero-point entropy ($S_\textrm{\,imp}(0)$), specific heat ($\Delta C$) and susceptibility ($\Delta\chi$) as the four-channel Kondo model, and it has a coupling strength-dependent, non-universal Wilson ratio ($R_W$),
\bea
S_\textrm{\,imp}(0)=\log(\sqrt{3}),\,\,\Delta C\propto T^{\,2/3},\,\, \Delta\chi \propto T^{\,-1/3}.
\eea
The results for the zero-point entropy and the $T$-dependence of  impurity contribution to the entropy ($\Delta S$) and $\Delta\chi$ 
were reproduced with NRG in Ref.\ \cite{Galpin14}.
Exploiting the spherical symmetry and particle number conservation of the model with NRG \cite{Toth08}, we also reproduced these results, as well as the leading $T$-dependence of $\Delta C$ and the specific heat coefficient ($\gamma$), and the non-universality of $R_W$. Our results are presented in Fig.\ \ref{fig:1.5CK_TD}. Calculating diverging, thermodynamical properties, especially the specific heat coefficient  of NFL quantum impurity models with NRG is a notoriously difficult numerical task as evidenced by the scarcity of such results. Even when employing $z$-averaging \cite{Oliveira94} and using as many Abelian and non-Abelian symmetries of the system as possible \cite{Toth08}, the number of multiplets kept needs to be increased until the analytically predicted asymptotics are more or less reproduced. So NRG does not have much predictive power for diverging TD, despite the fact that the low-energy, finite-size excitation spectrum is easily calculated with high precision. 

For the experimental realization of the corresponding 1.5CK NFL behavior, based on the derivation of the model,
we looked for dilute heavy fermion compounds with cubic point symmetry at the $f$-element site and where the spin degeneracy of $\Gamma_{4/5}$ conduction electrons is lifted while the impurity retains its two-fold, Kramers degenerate ground state below $T_K$. This way
1.5CK physics might be achieved in cubic itinerant ferromagnets with sites that can be substituted with
impurities with $f$-valence electrons in Kramers doublet grounds states like a single $f$-electron as in Ce$^{3+}$ or
Pr$^{4+}$ or a single hole as in Tm$^{2+}$ ions. We identified three candidate systems, itinerant ferromagnetic
hosts with rare-earth ions as the impurity, fitting the above symmetry requirements. Our candidates
are ZrZn$_2$ substituted with Pr for Zr, CoS$_2$ substituted with Tm for Co \cite{Barakat05}, and YFe$_2$ substituted with Ce for Y \cite{Paolasini00}
at the level of approximately 1-3 at.$\%$. They can be viable 1.5CK systems, if the magnetic field at the
impurity sites generated by itinerant ferromagnetism can be kept below the Kondo scale, or if we can
offset this magnetic field acting on the impurity by an applied magnetic field similarly to the case of the
Jaccarino--Peter compensation effect \cite{Meul84}.
\subsection{Properties of the topological Kondo model under time-reversal} If the $\Gamma_{4/5}$ conduction electrons derive from
electrons whose spin degeneracies are lifted, and thus the $c$ operators are to be complemented with an extra spin index, then Eqs.\ \eqref{eq:top_Kondo1}-\eqref{eq:top_Kondo2} break time-reversal symmetry. Otherwise, without an extra spin index, the topological Kondo model is invariant under time-reversal as it is the scalar product of the impurity spin operator with the generalized spin or angular momentum operator of the conduction electrons.

\section{Kramers doublet impurity coupled to $\Gamma_8$ conduction electrons}
The final case to be discussed is when the impurity is a $\Gamma_{6}$ irrep \footnote{Working with a $\Gamma_{7}$ impurity gives identical results.} which is coupled to $\Gamma_8$ conduction electrons.  From the tensor product decompositions, Eqs.\ (\ref{eq:G8xG8}) and (\ref{eq:G6xG6}), we see that apart from the potential scattering, there are two possible couplings in the exchange Hamiltonian. One of them can be identified with the scalar product of the spin operator for a spin-half impurity and the local spin density of spin-$\frac 3 2$ conduction electrons, 
\bea
          &&{\cal  H}_{\,\textrm{spin-}1/2\cdot\textrm{spin-}3/2}^{\,\Gamma_4\otimes\Gamma_4}\,=\,{\cal J}{\vec S}^{\textrm{\,imp}}\,{\vec s}^{\textrm{\,c}}\nn\\
          &&\equiv\,{\cal J}_\perp\left({S}^{\textrm{\,imp}}_x\,{s}^{\textrm{\,c}}_x+{S}^{\textrm{\,imp}}_y\,{s}^{\textrm{\,c}}_y\right)\,+\,{\cal J}_z{S}^{\textrm{\,imp}}_z\,{s}^{\textrm{\,c}}_z\label{eq:spin-half-spin-3/2-Kondo}
\eea
where ${\cal J}_\perp={\cal J}_z\equiv{\cal J}$, the $x,y,z$ components of ${\vec s}^{\textrm{\,c}}$ are given by
\begin{widetext}
\bea
{\vec s}^{\textrm{\,c}}&=&\left[
\begin{array}{c}
  \left(\sqrt{\frac{3}{4}}\,\Psi_{\Gamma_8}^{-\frac 3 2\,\dagger}\,\Psi_{\Gamma_8}^{-\frac 1 2}\,+\,\Psi_{\Gamma_8}^{-\frac 1 2\,\dagger}\,\Psi_{\Gamma_8}^{\frac 1 2}\,+\,\sqrt{\frac{3}{4}}\,\Psi_{\Gamma_8}^{-\frac 1 2\,\dagger}\,\Psi_{\Gamma_8}^{-\frac 3 2}\,
+\,\sqrt{\frac{3}{4}}\,\Psi_{\Gamma_8}^{\frac 1 2\,\dagger}\,\Psi_{\Gamma_8}^{\frac 3 2}\,+\,\Psi_{\Gamma_8}^{\frac 1 2\,\dagger}\,\Psi_{\Gamma_8}^{-\frac 1 2}\,+\,\sqrt{\frac{3}{4}}\,\Psi_{\Gamma_8}^{\frac 3 2\,\dagger}\,\Psi_{\Gamma_8}^{\frac 1 2}\,\right)\\
  i\left(\sqrt{\frac{3}{4}}\,\Psi_{\Gamma_8}^{-\frac 3 2\,\dagger}\,\Psi_{\Gamma_8}^{-\frac 1 2}\,+\,\Psi_{\Gamma_8}^{-\frac 1 2\,\dagger}\,\Psi_{\Gamma_8}^{\frac 1 2}\,-\,\sqrt{\frac{3}{4}}\,\Psi_{\Gamma_8}^{-\frac 1 2\,\dagger}\,\Psi_{\Gamma_8}^{-\frac 3 2}\,
+\,\sqrt{\frac{3}{4}}\,\Psi_{\Gamma_8}^{\frac 1 2\,\dagger}\,\Psi_{\Gamma_8}^{\frac 3 2}\,-\,\Psi_{\Gamma_8}^{\frac 1 2\,\dagger}\,\Psi_{\Gamma_8}^{-\frac 1 2}\,-\,\sqrt{\frac{3}{4}}\,\Psi_{\Gamma_8}^{\frac 3 2\,\dagger}\,\Psi_{\Gamma_8}^{\frac 1 2}\,\right)\\
  \left(-\frac{3}{2}\,\Psi_{\Gamma_8}^{-\frac 3 2\,\dagger}\,\Psi_{\Gamma_8}^{-\frac 3 2}\,-\,\frac 1 2\,\Psi_{\Gamma_8}^{-\frac 1 2\,\dagger}\,\Psi_{\Gamma_8}^{-\frac 1 2}\,+\,\frac{1}{2}\,\Psi_{\Gamma_8}^{\frac 1 2\,\dagger}\,\Psi_{\Gamma_8}^{\frac 1 2}\,
+\,\frac{3}{2}\,\Psi_{\Gamma_8}^{\frac 3 2\,\dagger}\,\Psi_{\Gamma_8}^{\frac 3 2}\,\right)
\end{array}
\right]\,
\eea
\end{widetext}
and ${\vec S}^{\textrm{\,imp}}$ is defined in Eq.\ \eqref{eq:spin_def}.
${\cal  H}_{\,\textrm{spin-}1/2\cdot\textrm{spin-}3/2}^{\Gamma_4\otimes\Gamma_4}$ is a cubic invariant only if ${\cal J}_\perp={\cal J}_z$, in which case it also has spherical symmetry. For arbitrary real couplings is has time-reversal symmetry. Even though cubic symmetry sets ${\cal J}_\perp={\cal J}_z$, in the second line of Eq.\ \eqref{eq:spin-half-spin-3/2-Kondo}, we allowed for ${\cal J}_\perp\neq{\cal J}_z$, to point out that according to our NRG calculations, for ${\cal J}_\perp < {\cal J}_z$, this model is a Fermi liquid, whereas for ${\cal J}_\perp > {\cal J}_z$, it flows to the 2CK fixed point. Yet for ${\cal J}_\perp = {\cal J}_z = {\cal J}$, this model has a different kind of NFL fixed point and was solved using CFT and NRG in Ref.\ \cite{Kim97}. We reproduced the NFL, finite-size energy spectrum with NRG making use of the spin symmetry and the particle number conservation of the model. For ${\cal J}_\perp = {\cal J}_z$, ${\cal  H}_{\,\textrm{spin-}1/2\cdot\textrm{spin-}3/2}^{\Gamma_4\otimes\Gamma_4}$ is claimed to have identical properties in the spin sector to the ten-channel Kondo model \cite{Fabrizio96}. According to the CFT calculations, its thermodynamics is described by the following  relations
\bea
&&S_\textrm{imp}(0)=\log\left(\frac{2}{\sqrt{6}-\sqrt{2}}\right)\approx\log(1.932)\,,\nn\\
&&\Delta C\propto T^{\,1/3},\quad \Delta\chi \propto T^{\,-2/3}\,.
\eea
The question whether this type of NFL behavior is symmetry protected can be decided by solving the second allowed exchange coupling in this setup. The exact form of this coupling is lengthy and is relegated to the App.\ \ref{app:G6_G8}. This coupling does not have spherical symmetry only time-reversal and cubic symmetry, but as it turns out after solving it with NRG, spherical symmetry is recovered as part of conformal symmetry around the $T=0$ NFL fixed point as this model flows to the same fixed point as Eq.\ \eqref{eq:spin-half-spin-3/2-Kondo} as well as the sum of the two couplings. Thus this type of NFL behavior is guaranteed by cubic symmetry to dominate at low $T$.
It could be achieved in cubic, nonmagnetic metals with sites that can be substituted with
impurities with $f$-valence electrons in Kramers doublet grounds states like a single $f$-electron or a single hole, and where band electrons of $\Gamma_8$ symmetry screen the impurity. 

\section{Conclusion}
We derived every cubic symmetry-allowed NFL exchange coupling between an impurity ion with a doublet ground state and local conduction electrons that does not require accidental degeneracy for its realization, and found three different NFL quantum impurity behaviors that can arise in this setup. In the process, we obtained a novel cubic exchange coupling which flows to the same fixed point as the spin-half impurity spin-$\frac 3 2$ conduction electron Kondo model. We found that this relatively little-studied critical behavior has the greatest chance of existing in cubic systems by showing that it is protected by cubic symmetry in contrast to the 2CK Kondo behavior that was suggested by Cox in 1987 \cite{Cox87} but proved difficult to realize in real materials. We pointed out that Cox's setup \cite{Cox87} is not guaranteed to lead to 2CK physics since cubic symmetry does not prevent an effective spatial anisotropy from exceeding the 2CK coupling, which restores a Fermi liquid behavior, and that the  quadrupolar Kondo model breaks time-reversal symmetry by construction. We derived the topological Kondo model in cubic symmetry and showed that a Kramers doublet impurity coupled to a conduction electron triplet gives a NFL. Although we used cubic symmetry, we argued that our treatment is the most general as far as doublet impurities are concerned  and thus no further types of NFL behavior are expected to arise in different crystal field settings and without accidental degeneracy. We also computed for the first time the temperature dependence of the specific heat coefficient and the Wilson ratio  in the topological Kondo model with NRG for the full temperature range by exploiting the SU(2)$_\textrm{spin}\times$U(1)$_\textrm{charge}$ symmetry of the system and employing $z$-averaging, and validated previous analytic results that apply only in the NFL scaling regime around $T=0$.
          
\section{Acknowledgments} 
This project has received funding from the
European Union’s Horizon 2020 research and innovation programme under the Marie Skłodowska-Curie grant agreement No 101024548.
A.T.\ thanks Sz\'ofia M\'atrai for her language-related help, and acknowledges support from the Daphne Jackson Trust.

\appendix

\section{Kondo type of exchange couplings in cubic symmetry for a $\Gamma_3$ ($E$) non-Kramers doublet impurity and $\Gamma_5$ ($T_2$) conduction electrons}
%\section{$\Gamma_3$ impurity coupled to $\Gamma_5$ conduction electrons}
\label{app:G3_G5}

The impurity ground state doublet is given by the ket vectors $\,|\Gamma_{3}^{\,3z^2-r^2}\rangle$ and $\,|\Gamma_{3}^{\sqrt{3}(x^2-y^2)}\rangle$. They are combined according to
$\Gamma_3\,\otimes\,\Gamma_3\,=\,\Gamma_1\,\oplus\,\Gamma_2\,\oplus\,\Gamma_3\,$ \cite{Koster63}. As for the product of the two local, $\Gamma_5$
conduction electron and hole tensor operators,
$\,
\left(
\begin{array}{c}
c^\dagger_{yz}\\
c^\dagger_{zx}\\
c^\dagger_{xy}\\
\end{array}
\right)$ and
$\left(
\begin{array}{c}
c_{yz}^{}\\
c_{zx}^{}\\
c_{xy}^{}
\end{array}
\right),\,$
whose spin degeneracies are lifted, they can be decomposed according to 
$\Gamma_5\,\otimes\,\Gamma_5\,=\,\Gamma_1\,\oplus\,\Gamma_3\,\oplus\,\Gamma_4\,\oplus\,\Gamma_5$. Thus the exchange Hamiltonian for a $\Gamma_3$
impurity ground state hybridizing with $\Gamma_5$ conduction electrons has two independent couplings. 
Apart from the ubiquitous and exactly marginal potential scattering term, i.e.\
\begin{widetext}
\bea
{\cal J}^{\Gamma_1\otimes\Gamma_1}_{\Gamma_3\,\textrm{imp},\Gamma_5\,\textrm{cond}\,e^-}\left(|\Gamma_3^{\sqrt{3}(x^2-y^2)}\rangle\langle \Gamma_3^{\sqrt{3}(x^2-y^2)}|+|\Gamma_3^{3z^2-r^2}\rangle\langle \Gamma_3^{3z^2-r^2}|\right)
\,\left(c^\dagger_{yz}c_{yz}^{}\,+\,c^\dagger_{zx}c_{zx}^{}\,+\,c^\dagger_{xy}c_{xy}^{}\right)
\eea
which is already diagonal in both the impurity
and the conduction electron degrees of freedom,
there is only one cubic symmetry-allowed term in the exchange Hamiltonian, namely
\bea
 {\cal J}^{\Gamma_3\otimes\Gamma_3}_{\Gamma_3\,\textrm{imp},\Gamma_5\,\textrm{cond}\,e^-}\left[\frac{1}{\sqrt{3}}\left(|\Gamma_3^{3z^2-r^2}\rangle\langle \Gamma_3^{3z^2-r^2}|-|\Gamma_3^{\sqrt{3}(x^2-y^2)}\rangle\langle \Gamma_3^{\sqrt{3}(x^2-y^2)}|\right)\left(c^\dagger_{yz}c_{yz}^{}\,+\,c^\dagger_{zx}c_{zx}^{}\,-\,2c^\dagger_{xy}c_{xy}^{}\right)\right.\nn\\
      \left.\left(|\Gamma_3^{3z^2-r^2}\rangle\langle \Gamma_3^{\sqrt{3}(x^2-y^2)}|\,+\,|\Gamma_3^{\sqrt{3}(x^2-y^2)}\rangle\langle \Gamma_3^{3z^2-r^2}|\right)\left(c^\dagger_{yz}c_{yz}^{}\,-\,c^\dagger_{zx}c_{zx}^{}\right)\right]\,.
 \eea
 \end{widetext}
It too is almost already diagonal, only the impurity part needs to be diagonalized which can be done independently of the conduction electrons.

\section{Kondo type of exchange couplings in cubic symmetry\\for a Karmers doublet ($\Gamma_6$ or $\Gamma_7$) impurity and $\Gamma_8$ conduction electrons}
\label{app:G6_G8}
Now we choose the impurity to be in a $\Gamma_6$ state. As the Clebsch--Gordan coefficients for $\,\Gamma_6\,\otimes\,\Gamma_6\,$ and $\,\Gamma_7\,\otimes\,\Gamma_7\,$ are identical \cite{Koster63}, the form of the exchange Hamiltonians do not depend on this choice. The basis functions of the $\Gamma_6$ irreducible representation (of the point groups $O$ or $T_d$) can be chosen  
as the tensor operator doublets 
$\,
\left[
\begin{array}{c}
f^\dagger_\uparrow\\
f^\dagger_\downarrow
\end{array}
\right],\,
\left[
\begin{array}{c}
\,f^{}_\downarrow\\
-f^{}_\uparrow
\end{array}
\right],\,$
creating and annihilating spin-half electrons. They can be combined according to the rule
$\Gamma_6\,\otimes\,\Gamma_6\,=\,\Gamma_1\,\oplus\,\Gamma_4$  to get the charge (i.e.\ particle number) conserving operators:
$\propto\left(f^\dagger_\uparrow\,f^{}_\uparrow\,+\,f^\dagger_\downarrow\,f^{}_\downarrow\right)$ transforming as a scalar (or $\Gamma_1$), and
$\propto\,\left[
\begin{array}{c}
  \left(f^\dagger_\uparrow\,f^{}_\downarrow\,+\,f^\dagger_\downarrow\,f^{}_\uparrow\right)\\
  -i\left(f^\dagger_\uparrow\,f^{}_\downarrow\,-\,f^\dagger_\downarrow\,f^{}_\uparrow\right)\\
  \left(f^\dagger_\uparrow\,f^{}_\uparrow\,-\,f^\dagger_\downarrow\,f^{}_\downarrow\right)
\end{array}
\right]\,$ transforming as the $x,y,z$ components of ${\vec S}^{\textrm{\,imp}}$ i.e.\ as a $\Gamma_4$ irrep. The basis functions of the $\Gamma_8$ irreducible double-valued
representation (i.e.\ spinor) can be chosen to be the
quartets creating/annihilating spin-$\frac 3 2$ fermions
$\,
\left[
\begin{array}{c}
c^\dagger_{\frac 3 2}\\
c^\dagger_{\frac 1 2}\\
c^\dagger_{-\frac 1 2}\\
c^\dagger_{-\frac 3 2}
\end{array}
\right],\,
\left[
\begin{array}{c}
c^{}_{-\frac 3 2}\\
-\,c^{}_{-\frac 1 2}\\
c^{}_{\frac 1 2}\\
-\,c^{}_{\frac 3 2}
\end{array}
\right].\,$
Their product can be decomposed into the following irreps 
$\Gamma_8\,\otimes\,\Gamma_8\,=\Gamma_1\,\oplus\,\Gamma_2\,\oplus\,\Gamma_3\,\oplus\,2\Gamma_4\,\oplus\,2\Gamma_5$. There are two independent particle-hole irreducible tensor operators that transform as $\Gamma_4$ triplets. The first one is %anti-Hermitian, we make it Hermitian by mutliplying it with $-i$:
\begin{widetext}
\bea
\left[
\begin{array}{c}
  \left(\sqrt{\frac{3}{4}}\,c^\dagger_{-\frac 3 2}\,c^{}_{-\frac 1 2}\,+\,c^\dagger_{-\frac 1 2}\,c^{}_{\frac 1 2}\,+\,\sqrt{\frac{3}{4}}\,c^\dagger_{-\frac 1 2}\,c^{}_{-\frac 3 2}\,
+\,\sqrt{\frac{3}{4}}\,c^\dagger_{\frac 1 2}\,c^{}_{\frac 3 2}\,+\,c^\dagger_{\frac 1 2}\,c^{}_{-\frac 1 2}\,+\,\sqrt{\frac{3}{4}}\,c^\dagger_{\frac 3 2}\,c^{}_{\frac 1 2}\,\right)\\
  i\left(\sqrt{\frac{3}{4}}\,c^\dagger_{-\frac 3 2}\,c^{}_{-\frac 1 2}\,+\,c^\dagger_{-\frac 1 2}\,c^{}_{\frac 1 2}\,-\,\sqrt{\frac{3}{4}}\,c^\dagger_{-\frac 1 2}\,c^{}_{-\frac 3 2}\,
+\,\sqrt{\frac{3}{4}}\,c^\dagger_{\frac 1 2}\,c^{}_{\frac 3 2}\,-\,c^\dagger_{\frac 1 2}\,c^{}_{-\frac 1 2}\,-\,\sqrt{\frac{3}{4}}\,c^\dagger_{\frac 3 2}\,c^{}_{\frac 1 2}\,\right)\\
  \left(-\frac{3}{2}\,c^\dagger_{-\frac 3 2}\,c^{}_{-\frac 3 2}\,-\,\frac 1 2\,c^\dagger_{-\frac 1 2}\,c^{}_{-\frac 1 2}\,+\,\frac{1}{2}\,c^\dagger_{\frac 1 2}\,c^{}_{\frac 1 2}\,
+\,\frac{3}{2}\,c^\dagger_{\frac 3 2}\,c^{}_{\frac 3 2}\,\right)
\end{array}
\right]
\eea
\end{widetext}
that correspond to the $x,y,z$ components of ${\vec s}^{\textrm{\,c}}$, the local spin density of spin-$\frac 3 2$  conduction electrons.
From this tensor operator multiplet and ${\vec S}^{\textrm{\,imp}}$, the following exchange Hamiltonian can be constructed upon taking their scalar product
\bea\label{appeq:spin-half-spin-3/2-Kondo}
{\cal  H}^{\textrm{\,spin-}1/2\cdot\textrm{spin-}3/2}_{\Gamma_4\otimes\Gamma_4}&=&{\cal J}{\vec S}^{\textrm{\,imp}}\,{\vec s}^{\textrm{\,c}}\,,
\eea
which has a non-Fermi liquid fixed point \cite{Kim97}.
The second $\Gamma_4$ conduction-electron-hole irreducible tensor operator %one also comes out as anti-Hermitian and we multiply it with $-i$ as well\\
\begin{widetext}
\bea
\propto\left[
\begin{array}{c}
  \left( 5\,c^\dagger_{-\frac 3 2}\,c^{}_{\frac 3 2}\,-\,\sqrt{3}\,c^\dagger_{-\frac 3 2}\,c^{}_{-\frac 1 2}\,+\,3\,
  c^\dagger_{-\frac 1 2}\,c^{}_{\frac 1 2}\,-\,\sqrt{3}\,c^\dagger_{-\frac 1 2}\,c^{}_{-\frac 3 2}\,-\,\sqrt{3}\,c^\dagger_{\frac 1 2}\,c^{}_{\frac 3 2}\,
  +\,3\,c^\dagger_{\frac 1 2}\,c^{}_{-\frac 1 2}\,-\,\sqrt{3}\,c^\dagger_{\frac 3 2}\,c^{}_{\frac 1 2}\,
  +\,5\,c^\dagger_{\frac 3 2}\,c^{}_{-\frac 3 2}\,\right)\\
  i\left(-\,5\,c^\dagger_{-\frac 3 2}\,c^{}_{\frac 3 2}\,-\,\sqrt{3}\,c^\dagger_{-\frac 3 2}\,c^{}_{-\frac 1 2}\,+\,3\,
  c^\dagger_{-\frac 1 2}\,c^{}_{\frac 1 2}\,+\,\sqrt{3}\,c^\dagger_{-\frac 1 2}\,c^{}_{-\frac 3 2}\,-\,\sqrt{ 3 }\,c^\dagger_{\frac 1 2}\,c^{}_{\frac 3 2}\,
  -\,3\,c^\dagger_{\frac 1 2}\,c^{}_{-\frac 1 2}\,+\,\sqrt{3}\,c^\dagger_{\frac 3 2}\,c^{}_{\frac 1 2}\,
  +\,5\,c^\dagger_{\frac 3 2}\,c^{}_{-\frac 3 2}\,\right)\\
  -\,2\,\left(\,c^\dagger_{-\frac 3 2}\,c^{}_{-\frac 3 2}\,-\,3\,c^\dagger_{-\frac 1 2}\,c^{}_{-\frac 1 2}\,+\,3\,c^\dagger_{\frac 1 2}\,c^{}_{\frac 1 2}\,
-\,c^\dagger_{\frac 3 2}\,c^{}_{\frac 3 2}\,\right)
\end{array}
\right]\nn\\
\eea
\end{widetext}
can also be used to build a scalar operator by taking its scalar product with ${\vec S}^{\textrm{\,imp}}$. Note that the terms containing $c^\dagger_{-\frac 3 2}\,c^{}_{\frac 3 2}$ and $c^\dagger_{\frac 3 2}\,c^{}_{-\frac 3 2}$ break spherical symmetry. Upon solving the resulting impurity Hamiltonian with the
numerical renormalization group, it turns out that it flows to the same fixed point as Eq.\ \eqref{appeq:spin-half-spin-3/2-Kondo}. Thus, in this setup, all possible exchange couplings, apart from the potential scattering, flow to the same NFL fixed point, thus this fixed point is protected in cubic symmetry contrary to the two-channel Kondo fixed point for a $\Gamma_3$ impurity ground state coupled to $\Gamma_8$ conduction electrons as discussed in the main part of the paper.

%\vspace{-2.2em}

\end{document}